THE SEMIOTIC MACHINE

Eric Engle

*"Humans are no longer the only and lonely ones that are capable of, and to some degree dependant on, the interpretation of signs."*                    Frieder Nake

I. Introduction

A. Problématique

This article attempts to answer the question: What can semiotics of the algorithmic sign tell us about the form and substance of interactive algorithmic art?

I present two answers to the question:
1) Semiotics can be used to analyze the user interface we create to represent interactive algorithmic art (form).
2) Semiotics can also be used to analyze algorithmic art (substance).
However this thesis will focus primarily on demonstrating an example of semiotic analysis of algorithmic art and only secondarily on an analysis of interface design. Thus the thesis of this paper is that semiotics can be used to analyze algorithmic art. A semiotic analysis of algorithmic art revelas that algorithmic art represents a synthesis of classical and early modern rationalism with late-modern post-structuralism via a dialectical synthesis of rule making (algorithm) and rule breaking (art).

To answer the question "What can semiotics of the algorithmic sign tell us about the form and substance of interactive algorithmic art?" several preliminary questions must also be answered. These are:

What is semiotics?
What is the algorithmic sign?
What is interactive algorithmic art?
What is algorithmic art?
What is interactive art?

Additionally, the article provides numerous basic terms and definitions in semiotics. These definitions are often necessary in order to answer the problématique proposed and are also provided to establish "common ground" for the work.

B. Definitions and Common Terms

1. Semiotics

Semiosis is the science of signs and their meanings. While the semiotics of computer programs is a new field, looking at software as a set of signs makes sense.

"On a descriptive level, semiotics is a powerful theory. As long as we do not expect a predictive potentional from semiotics, we can use it nicely to describe certain differences between communicative processes." (Nake, 2005.)

That is, we can use semiotics to describe software. But, we cannot expect semiotics to tell us much about future developments in software. Semiotics basically looks at past and present definitions of the relations among signs and our interpretations of them. To the extent that semiotics deals with arbitrary signs (symbols) it cannot on its own terms be predictive but rather would have to resort to other sciences such as economics or history for predictive power.

2. Interactive art

Algorithmic art is not necessarily interactive art and vice verse.
Interactive art 1) Involves the user and 2) Invokes emotions in the user. It calls upon the users sense of play to engage them in the work.

Interactive art is art in which "the viewer must do something in order to make something out of it or have an experience... Every interactive work tries to entice its user to perform certain actions, in the hope that the work will thereby reorganize itself into an unforeseen coherence.

Interactive art is by definition non-autonomous and organizes itself as an open system, which functions via the exchange of matter, energy and/or information with the environment (i.e. the visitor)."
(DEAF: 2004); Interactive art "involves the spectator in some way. Some sculptures achieve this by letting the observer walk in, on, and around the piece. Other works include computers and sensors to respond to motion, heat or other types of input. Many pieces of Internet art and electronic art are highly interactive. Sometimes visitors are able to navigate through a hypertext environment; some works accept textual or visual input from outside; sometimes an audience can influence the course of a performance or can even participate in it." (Wikipedia, Interactive Art: 2005). Interactive art is intended, generally, not as an attack on the audience but expects the audience to be aware of media generally (Feingold, 2005).

3. Sign

The term sign is used by both Saussure and Peirce but with rather different meanings. Thus we must explain what both meant by this term.

    a. Saussure
        i. A Structural Theory of the Sign

Saussure, as a structuralist. believed that (only) by studying and understanding the structures that frame and form social action can we understand any social phenomenon.  For Saussure, the sign is a composite: the sign is constituted by a signifier (a sound or image) and a signified (an object associated with the sound or image).

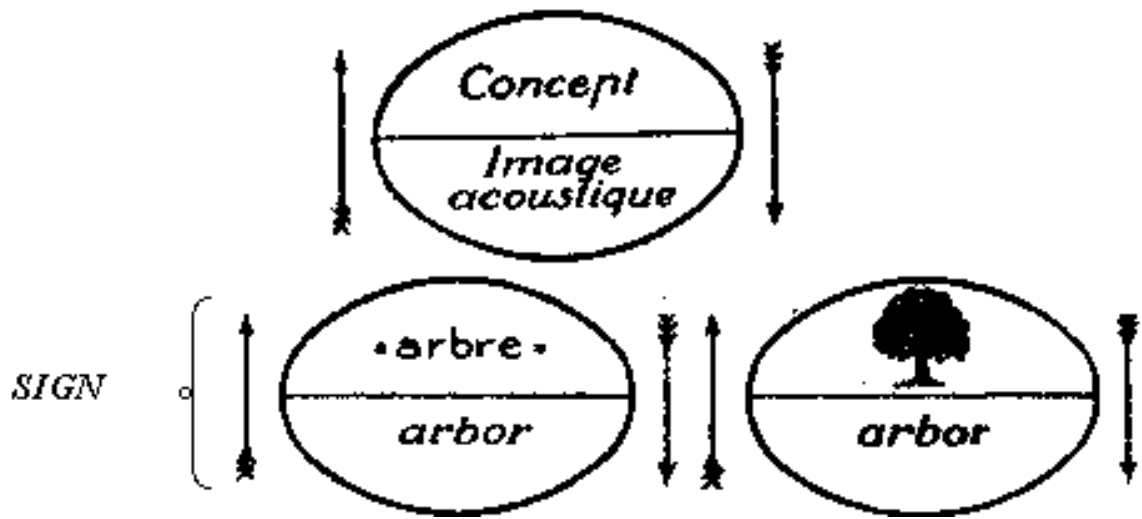

        ii. A linguistic theory of the sign

As well as being structural, Saussure's approach is linguistic: he sees the sign as part of language. Since we cannot understand parts of a thing unless we understand the whole thing we cannot
understand the sign in isolation but rather as a part of a language. However, he looks at the language as it is: not the processes by which languages evolve.

Since only humans use language Saussure's sign is anthropocentric. Saussure did not discuss the use of signs among animals nor non-linguistic signs. For Saussure, language is a system of signs.

Saussure's theory is also a theory of the sign in the mind, not in action. The sign has signification because of our subjective understanding of it. The sign to Saussure is arbitrary. This was not the view of pre-scientific thought!

iii. The Arbitrary Sign?

For Saussure the sign was completely arbitrary. This was not however the view of antiquity: neither Hebrew nor Sanskrit saw the sign as arbitrary. Quite the opposite, the sign was seen as eternally true, through time and space as a manifestation of the divine.

More realistically, we know some sounds, like 'buzz' or 'click', mimic the sound of that which they represent. Further, it seems likely that all human languages have one ultimate common root since some words, like 'ma' for mother, are just about universal. Still the linguistic sign is seen as "arbitrary in the sense that there is no relation between the sound of a word and its meaning other than convention, a 'contract' or rule." [Iverson, 1986: p. 85.] In contrast "visual signs are not arbitrary but 'motivated' - there is some rationale for the choice of signifier." [Iverson, 1986: p. 85.]  Artistic representation is a reflection of the object represented even oftener than spoken representation:
"Saussure's stress on arbitrary signs and his insight into how they rely on a systematic play of both conceptual and phonetic differences can inform our understanding of visual signs. Visual signs may be motivated and yet still obey some of the semiotic principles most clearly realizued by language. The Saussurean model of semiotics is in fact a valuable antidote to the lingering assumption that the relation between the visual sign and its object is a natural and immediate one." [Iverson, 1986: p. 85.]

b. Peirce

Unlike Saussure, who defines the sign statically, Peirce situates the sign in its social context of dynamic evolution.

Peirce defines the sign as consisting of a representamen (the sign itself) an interpretant (the idea the sign points to) and the ground (defining characteristics of the object) as well as an object. Peirce's terms are less than clear. For example, reconstructions of Peirce, for whatever reason, often omit the ground as a defining element of Peirce's sign.

For Peirce the Sign is a representamen - a pointer to something else. The representamen is the idea being communicated by the emitter of the sign. The interpretant is the sign created in the mind of an audience.

Peirce says that the representamen - itself a sign - is a pointer that points to other signs(!) (Later we will see we might be able to call the representamen an indexical sign). The other signs that the representamen points to, themselves signs, are known as the interpretant. They refer in turn to something Peirce calls the ground. The ground is the form of the material object. Logically speaking, the representamen must also refer to the ground.

We could call the representamen the signifier and the object the signified. I would even say we could call the interpretant a signifier as well. We might even call the ground a part of the signified. Thus the signifier is the representamen and interpretant and the signfied is the object and ground.

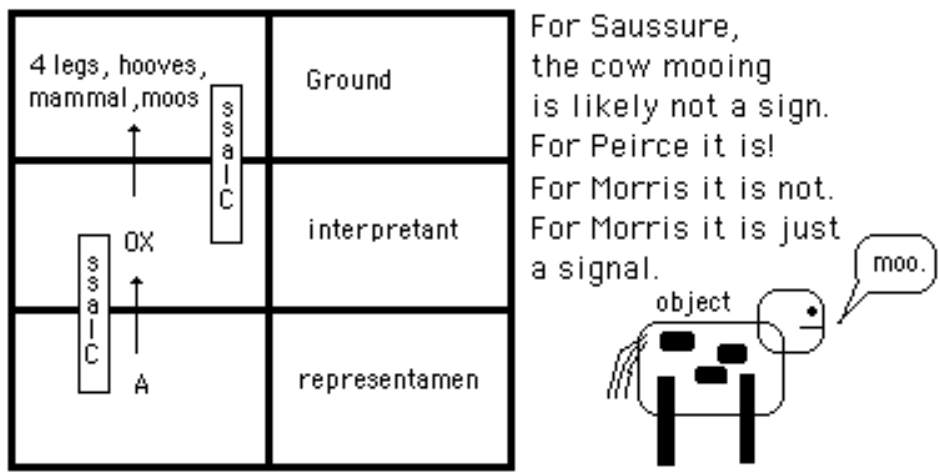

It is certain that the interpretant is a subjective interpretation of the representamen and by implication that the representamen is an objective representation of the signified object. If we try to look at Peirce's sign from Moriss's point of view then we might also say that the representamen is the sign on the syntactic level, that the ground is the sign on the pragmatic level and that the ground is the sign on the semantic level. The interpretant, as a subjective interpretation of the sign, must also be on the pragmatic level, though the interpretant also has a syntactic and semantic level.

Part of the problem in trying to look at Peirce using either Saussure or Morris is that Peirce's definition of the sign is autological. Both the interpretant and representamen are signs, yet they are also parts of signs. Logically, this means the sign is never completely defined: Each new sign we invoke implies a new representamen and interpretant which in turn each invoke: a new representamen and interpretant! And this ad infinitum! A truly recursive function will at some point have a test for termination. Peirce's sign contains no test for termination. Thus it is more exact to see the Peircian sign as autological as it is not, strictly speaking recursive since it is non-terminating. Alternatively we could simply reject or redefine Peirce.

Some argue that the Peircian sign is recursive. While it might be useful to define signs recursively, Peirce does not in fact do so. Recursive functions have a test for satisfaction. Peirces definition of the sign cannot terminate. Thus Peirce's sign is not in fact defined recursively though his definition of the sign is autological.

### Semiotics, Grammar and Rhetoric

Peirce appears to take up - and rename - the classical taxonomy of Aristotle. He describes and distinguishes between
Semiotics - which corresponds to Aristotle's Logic
Pure Grammar - which corresponds to grammar
Pure Rhetoric - which corresponds to rhetoric
This ontology makes the mistake of needlessly renaming known scientific objects. That adds an unecessary level to scientific discourse and results in potential confusion.

### c. Morris
#### 1.3 Morris

Morris was a behaviorist. For a behaviorist empirical reality is the only reality. Behaviorists study only observable phenomena. That is, behaviorism is the empirical study of stimuli and responses. Behaviorism ignores speculation as to the causes or hidden processes that generate observable phenomena. For Morris the sign directs behavior. It is "roughly: something that directs behavior with respect to something that is not at the moment a stimulus. (1946: 366). Morris wishes to develop an "empirical science of signs" (1946: 105-106) based on observation of stimulus & response. For Morris, the sign directs behavior.

Morris presents a social model of the sign. For Morris, the sign has no existence outside the community creating it. "something is a sign only because it is interpreted as a sign of something by some interpreter" (1938).

Morris defined Semiosis as a "sign process, that is a process in which something is a sign to some organism" (1946:366).

Morris's social view of semiology creates a problem however. If the sign is arbitrary (Saussure) and culturally determined (Morris, possibly Peirce as well) then no fixed permanent response can be observed in conjunction with a sign. Of course, temporally conditioned observations would be possible. But then Morris's theory of signs becomes, despite empiricism, not episteme but dialectic!

### Episteme and Dialectic

What is episteme? Episteme means knowledge, i.e. the object of science. Epistemic knowledge is invariable: For example, 'Every time A happens, B also happens' is an epistemic statement. Thus natural science is almost always epistemic.

What is dialectic? In dialectic, we do not have precise exact knowledge. Rather, by comparing the leading opinions of those well versed in a field we obtain the closest possible approximation to the truth.

Human sciences are generally dialectical. If semiology is dialectical, then why is Peirce fixated on empirical observation? In my opinion because he does not understand the distinction between episteme and dialectic.

### d. Eco
#### Communication

Synthesizing the different definitions of sign may be best obtained by placing semiotics in the context of communication as described by Umberto Eco. Eco defines communication as the transfer of a sign or signal. The source can be human or non-human (at least, if the non-human source follows rules understood by humans). So long as the signal or sign evokes an interpretive response in the addressee we have communication. Thus, a code is necessary for communication to occur. (Eco, 1976: pp. 8-9). This encoding is also seen initially in the work of Saussure, who also describes communication as a process of coding and decoding as illustrated below:

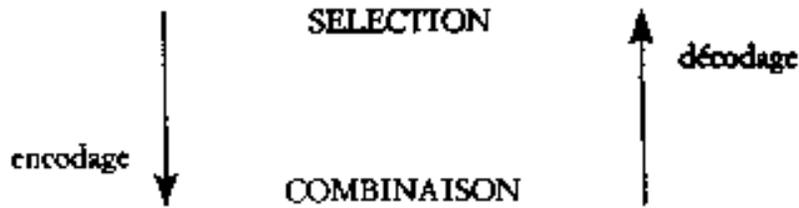

Codes

Thus for Eco codification is what defines a sign. Without codification we would have only a signal. The signal becomes a sign because it is an expression of a culturally codified meaning. For Eco, semiotics consists of two elements: a theory of codes and a theory of sign production. The theory of codes, structurally oriented, explains signification (substance). The theory of sign production, process oriented, explains communication phenomena (form). [de Souza, 1993: p. 763.] Because an artificial language system is an integral part of the computer computer science has an important linguistic component and sign production is central to computer science. [de Souza, 1993: p. 765.]

e. Synthesis

The sign is any linguistic or non-linguistic expression intended to communicate meaning by or to a person or animal. Saussure is the most coherent starting point of semiology but is not its end point. Semiology includes non-verbal communication, and communication of animals.

Semiosis is a process: language changes over time. At least some signs are arbitrary not only in the sense of being intermediaries but also in the sense of being conventional. If we understand the distinction between episteme and dialectic we will be able to understand and define the limits of behaviorism in semiology. But because the sign is contextualized socially it is to that extent a subject of dialectic and not of episteme. Behaviorism cannot address dialectical struggles because dialectical struggles, especially in their superstructural aspects, are purely mental entities and thus not part of science (episteme) as far as behaviorism is concerned all the more so because for the behaviorist only material observable phenomena are capable of scientific observation. The behaviorists simply ignore the possibility of dialectic developing well informed opinion (doxa) as a resolution of dialectical inquiry into social interactions.

Saussure's mental entity and Peirce's social process can be synthesized -- but not using Morris's behaviorism. Semiotics is in all likelihood a subject only of dialectic and not of episteme since signs do not always compel us to obey them and are, at least to some extent, arbitrary and conventional. This synthesis could occur as follows: We can see the SIGNIFIER as a synonym for the REPRESENTAMEN and the SIGNIFIED as a synonym for the OBJECT. Saussure does not discuss the ground, but it corresponds to Plato's concept of the form (eidos). This just leaves the INTERPRETANT to be defined. We know the interpretant requires an interpreter. So the interpretant is the subjective understanding of the sign in the mind of the person reading or hearing the sign. Both the interpretant and the representamen are signs, and the entire structure (interpretant, representamen, object and ground) is also a sign.

To avoid confusion I would use "signifiant" for signifier and "signifié" for signified when using Saussure's perspective. Saussure's definition of the sign is a definition - it terminates. Peirce's description of signs as a process is non-terminating. Thus it is not possible to exactly recast the sign as described by Peirce into Saussurian terms. However because the sign as described by Saussure could be recast to conform to Peirce their terminology can, at least to some extent, be synthesized, though imperfectly.

4. Icon, Index, Symbol

The multiple terms Peirce uses risk confusion. "Although the most popular of Peircian classifications is the icon-index-symbol one, there are dozens of others in his original proposal, many of which overlap, leading to a labyrinth of types and criteria" [deSouza, 1993: p. 761]. Worse, the terms index, icon and symbol have been taken up by computer science with meanings that are at times different from those used by Peirce. Thus we must define these terms as used by Peirce and Morris and as applied in computer science if only to avoid confusion but also possibly for application in a theory of the computable sign. We outright reject Morris, who appears self contradictory and define exactly what Peirce meant by the terms icon, index and symbol.

a. Icon

Icons are used in computer interfaces all the time. Popularly, the term means a pictorial representation of some metaphor or object, that is an abbreviation. However the term icon in semiotics means a particular type of pictogram. "An icon is a sign which refers to the object that it denotes merely by virtue of characters of its own and which it posesses". [Peirce, 1955,

p.102]; "An icon is something that looks like what it means" (Marcus, 1992: p. 52.) It "signifies by virtue of a similarity of qualities or resemblance to its object. For example, a portrait iconically repesents the sitter." [Iverson, 1986: p. 89]. The iconic sign signifies on the basis of "a relationship of similarity between the sign and its object" [Nöth 2002: p. 6] For example, an icon with a picture of a printer to represent the 'print' command is a true icon in the semiotic sense. In contrast, a button with a picture of an arrow is not an icon. That button would be an index. Iconic representation is independant of the object and interpretant. [Iverson, 1986: p. 89].

### b. Index

An index may be thought of as a pointer which signifies by reference. Indices are signs that are "'caused by the thing or process to which it refers... " (Marcus, 1992: p. 52.) For example, "a weathervane indexically signals the direction of the wind; a footprint indicates that someone has been on a beach." [Iverson, 1986: p. 89].
"Because the sign vehicle [of an index] is physically connected to its object, the interpreting mind has nothing to do with that connection except in noticing that it exists. Indexical signs don't depend on conventional codes to establish their meaning." [Iverson, 1986: p. 89.]
"Indexical signs... draw the interpreter's attention to their object by an immediate spatial, temporal or causal connection, are apparent in computer programming and text processing when the user is instructed by means of arrows, by the cursor or by commands such as assign, do, exit if or continue if...."[Nöth 2002: p. 6].

### c. Symbol

The symbol is a sign the signification of which is established by convention:
"The symbol signifies by virtue of a 'contract' or a rule. ... The tenuous, conventional relationship between sign-vehicle and object characteristic of the symbol relies upon an interpretant who knows the rule. To put it another way, there is an intrinsic dependance on the mind for there to be any relation at all. [Iverson, 1986: p. 89].

Because the symbol has signification due only to convention the symbol is arbitrary "A symbol is a sign that may be completely arbitrary in appearance." (Marcus, 1992: p. 52.) Words, at least in a western alphebet, and numbers are examples of symbols. [Nöth 2002: p. 6].

According to Morris, "[A] symbol is a sign produced by its interpreter which acts as a substitute for some other sign with which it is synonymous; all signs not symbols are signals." For Example: the heartbeat is a signal, but words describing it are a sign. (Morris, 1946:pp 100-101) cited in Nöth 1995: p. 50)"A symbol is a sign which refers to the object that it denotes by virtue of a law, usually an association of general ideas" [Peirce, 1955, p.102]: here, Morris contradicts himself. He says that a signal is a sign ("all signs not symbols are signals") when he later says "such signs are simply signals; his resulting words -- when substituted for such signals -- would however be symbols".

One avoids this contradiction simply by ignoring Morris who seems confused on the point whether a signal is or is not a sign as well as on behaviorism. Signals only become signs when interpreted as such. That is, a signal may have an interpretant but has no representamen. According to Peirce, there are three types of sign: icons, indexical signs, and symbols. This is the typology which will be used here, regardless of Morris.

### 5. Syntactics, Semantics, Pragmatics

Morris reiterates the error of Peirce in needlessly compounding intentional entites. This violates Occam's law. A scientific theory should use as few terms as possible to describe its object as completely as possible.

Morris describes the following fields as parts of semiotics: Syntactics, Semantics and Pragmatics. Morris, like Aristotle and Peirce, describes the same trivium, but uses different names for the same objects in a needless multiplication of intentional entities in violation of Occam's razor. To avoid confusion here are the terms used by Aristotle, Peirce and Morris to describe essentially the same phenomena:

| Aristotle | Peirce | Morris |
|---|---|---|
| Logic | Semiotics | Semantics |
| Grammar | Pure Grammar | Syntactics |
| Rhetoric | Pure Rhetoric | Pragmatics |

This trivium is not the only place where Aristotle influences Morris. Semiology, according to Morris, is both an instrument and object of science. That is exactly Aristotle's position as to logic: For Aristotle, logic is both a tool and object of science. How Morris could be so influenced by Aristotle on logic, grammar, and rhetoric without understanding the distinction between dialectic and episteme which voids his behaviorism escapes me.

Yet, precisely because these terms present a needless duplication of scientific effort and also because the terms "syntactics, semantics and pragmatics" have entered into scientific practice we must define them precisely.

### a. Syntactics

Syntactics is the processing of perceptual signals (colors, light, sound). "[S]yntactics - deals with combination of signs without regard for their specific significations or their relation to the behavior in which they occur." [Zemanek 1966: p. 139.]

### b. Semantics

Semantics is the assignation of meanings to perceptions and is culturally constructed. It is "the study of the relations of signs to the objects to which the signs are applicable ...semantics - deals with the signification of signs in all modes of signifying" [Zemanek 1966: p. 139.]

### c. Pragmatics

Pragmatics is the subjective individual meaning of sign, that is the interpretation of the sign in the mind of the individual (the interpretant) [Zemanek 1966: p. 139.] and "deals with the origin, uses and effects of signs within the behavior in which they occur [Zemanek 1966: p. 139.]. "'[P]ragmatics is the consideration of practical bearings of a notion' the syntactics and semantics of which are settled." [Zemanek 1966: p. 139.]

## II. Computational Semiotics

### A. The Semiotic Machine

#### 1. Software: A MetaCommunication Artefact

"The computer is at times called the semiotic machine." [Andersen, Nake, forthcoming: p. 7.]. Computer software is unique among human tools because it is interactive. Unlike other media, which do not react dynamically to the user's inputs, a computer program can generate different results based on different input provided by different users. For this reason a piece of software is semiotically unique. It is a metacommunication artefact - a message which sends and recieves messages: "systems ...are message senders and receivers at the immediate interface level, but they are also achieved messages, themselves, sent form designers to users thorugh the computational medium." [de Souza 1993: pp. 753-754]. Software represents "a highly peculiar type of message: a  message which sends and receives messages (a metacommunication artefact). [de Souza 1993: p. 773]

Computer programs are metacommunication artefacts not only because they are symbol processing machines [Nöth, 2002: p. 5]  but also because they are semiotic machines. "computers... operate with symbols... indexical and iconic signs (more precisely quasi signs...) [Nöth, 2002: p. 5] Thus we must explain the concept of the semiotic machine.

According to Frieder Nake, "The computer is at times called the semiotic machine."(Nake: 2005)"The semiotic machine is a machine to operate on signs." [Andersen, Nake forthcoming: p. 7]. Nöth agrees with this definition: "a semiotic machine... [is] a machine not restricted to the processing of symbols but also involved in other sign processes." [Nöth, 2002: p. 6].

To fully apprehend the idea of the computer as a semiotic machine we must explain the differences between each of the two communicants, the machine and the human. "Since programming languages are a communication link between man and amachine, we have two kinds of users: an artificial one and a natural one; a mechanized one and an illogical one. The first is fully algorithmic and carries out what the text means to him, while the second has notions, opinions and feelings. ..." This distinction explains our understanding of pragmatics - the computers and human pragmatics are different from each other. [Zemanek, 1966: p. 142] Thus, as we describe  human and computer semiotic interaction we will need to use or develop a concept of the computer sign, a quasi-sign, as part of a larger concept the algorithmic sign. The presence of these two very different communicants however raises cognitive dissonance: Humans communicate regularly - with other humans! They may occasionally communicate with animals. But they have only recently begun to communicate with machines. This schism between man and machine leads to a communicative discrepancy.

## 2. The Communicative Discrepancy

Though computers do not appear to have awareness, we tend to treat them as if they did. Nake observed this split and called it the communicative discrepancy of the computer. The communicative discrepancy is the result of anthropomorphism. [Anderson, Nake, forthcoming] Chris Crawford thinks it is inevitable and that software design should take advantage of it. Professor Nake is more cautious due to the confusion that anthropomorphism risks.

B. The Computer Sign

### 1. Signal, Computer Sign, Interpretation

Nake defines computable signals as a sign where the object is equivalent to the interpretant. The interpretant of a computer sign is calculated by a computable process of interpretation. [Andersen, Nake, forthcoming: p. 8.] Computer interpretation consists of "determining the one and only one meaning a sign has for the semiotic machine." [Andersen, Nake, forthcoming: p. 8.]

### 2. The Quasi Sign (a reduced sign)

The semiotic machine is interpreting a specific sign, the computable sign, which is a quasi-sign. "A quasi-sign is only in certain respects like a sign, but it does not fullfil all criteria of semiosis. While some criteria of semiosis may be present in machines, others are missing. The concept of quasi-sign thus suggests degrees of semioticiity. Quasi-semiosis... is thus the reduction ('degeneration' is Peirce's term) of a triadic sign process invlving a sign (representamen), affected by an object and creating an interpretant to a merely dyadic process with only a sign affected by its object." [Nöth, 2002: p. 8].

Computer semiosis is a quasi-semiosis because the signs involved are necessarily dyadic. Computational signs are necessarily dyadic because the computer "can only process signals... i.e. mechanical stimuli followed by automatic reactions." [Nöth, 2002: p. 9.]

### 3. Eidetic Nature of the Computer Sign

Computer signs, because they must be expressed as computable functions with only one interpretation also can be characterized as eidetic: Computer signs are similar to Platonic ideal entities:
"Many computer-based signs subscribe to a Platonic world-view in the sense that they possess an ideal, perfect version, and a material, imperfect shape. The perfect version works as a guideline for programmers building the material imperfect version. There are many examples: the imperfect graphical systems are based on ideal geometry, and the implemented semantics of programming languages are often based on a formal, logical specification. The dichotomy between purity and sin can be found in other areas, e.g. in linguistics, where competence is the idealized faculty of a language speaker, whereas performance describes his limitations of memory." [Andersen, Nake, forthcoming: pp. 10-11].

C. The Algorithmic Sign

If computers are semiotic machines then they are interpreting signs. The signs that computers interpret are represented algorithmically. Thus: "Software should be understood as an algorithmic sign. ...Algorithmic signs get interpreted twice: by a human and by a computer. ...Ontologically the computer is not interpreting. Logically it is. Therefore it is justified to define the algorithmic sign as indicated above. It is a sign between us and our computers."[Nake: 2005; Andersen Nake forthcoming p. 7.] So, algorithmic signs, according to Nake, have two interpreters, one human, and the other non-human (machine). They also have two existences, one directed inside the machine (machine readable) and one directed outside the machine (human readable):

> "Software objects in general and digital images in particular exist in two ways. They appear on the surface of the digital medium, and they are hidden away deep in its internal storage structures. The surface appearance is for our senses to perceive. The internal existence is for the computer processor to manipulate. The software object is a pair and unity of a perceiveable and a manipulable aspect. The two are inseperably interlinked. One does not exist without the other."[Nake: 2005] Algorithmic signs are stored in a stable form on media at the non-human input level. Yet at the level of human readable output they are dynamic and in need of constant maintenance and updating. [Andersen, Nake, forthcoming p. 10.] Nake also notes that this has been the history of signs generally, from the picture which led no double life, to the motion picture with its double life (stored and playing) - but still at least both levels were human readable. With television the media became no longer human readable in transmission. [Andersen, Nake, forthcoming p. 10.]

Another key feature of the algorithmic sign is its opacity. "algorithmic signs are normally not accessible to their users in their totality." [Andersen, Nake, Forthcoming: p. 11.] This is in fact very important for computer games. If the player knows the algorithm by which the game plays then the human will ordinarily win by exploiting the algorithm's weaknesses. With a transparent algorithm the only viable computer strategy, at least with current AI is a massive resources strategy since the limited information strategy would no longer be possible. Basically, the computer overwhelms the human by applying speed or numbers that make winning physically impossible for a human. However at least in some interactive algorithmic art installations it might be useful to have the algorithm, or at least parts of the algorithm, visible to the user. Of course, the machine code or assembler instructions would be of little use to the audience. However the source code or its natural language expressions in the user interface might be made explicit so that the user can better appreciate the subtle intricacies of algorithmic art. Thus the degree of invisibility of an algorithmic sign is another design choice and will depend on the message the author wishes to convey.

III. Art and Algorithm

A. Art Theory

   1. Art as a Sign

Semiotics is the science of signs and their interpretations. We know that signs may be words, even letters, as well as gestures or the spoken word. Artistic works - like computer programs - are also signs:

> "any message (a sequence or field of perceivable signals) contained
> information. The information content of a mesage could be measured.
> A painting could clearly be considered the carrier of signs. It could,
> indeed, be viewed as a complex sign composed of subsigns, which
> were in turn composed of subsigns and so on."(Nake: 2005)

This understanding of art as a sign leads to the question, how have artistic signs been interpreted historically? How are they interpreted today?

   2. Art in Antiquity and Early Modernity

In antiquity and even into early modernity artistic signs were seen as expressions of a rational process. Art in antiquity was seen as the medium of a either an orderly world reflected in its logical beauty (*kosmos* - art as logos in pre-modern theory) or of a world order imposed by the power of reason (*novus ordo seclorum* - art as reason's light or enlightened by reason in early modernity -- praxis), including the priveleged nature of writing versus image. But either form of rationalism could have been expressed algorithmically.

The fact that art privileged "the intellectual, abstractive procedures necessary for the production of an image, indicates that it is written against the background of a deep seated prejudice within western philosophy that iconic images are too close to their objects to have the character of thought and language. ...This is why most apologists for the visual arts, from the renaissance through the eighteenth century, have tended to valorize design, the intellectual conception of a work, as the invisible 'soul' of painting as opposed to its visible part, its 'flesh and body and have stressed analogies between painting and poetry or geometry or rhetoric."

[Iverson, 1986: pp 85-86]. Does algorithmic art feed this view of art? Art as ratio? I think so, at least potentially.

### 3. Art in Late Modernity

Late modernity (erroneously called 'post' modernity) rightly questions the paradigm of art in antiquity and early modernity, including the priveleged nature of writing versus image.
The late modern (better known as post modern) view grappled with the pre-modern view first using structuralism, then post-structuralism:
"The new approaches to art history are new, and share a common pursuit, to the extent that they begin with a radically different conception of art. Art is no longer regarded as part of the solution but as part of the problem, laden as it is with all the ideological baggage of history, be it bourgeois, racist or patriarchal. The new critical procedure, accordingly, involves a thorough going critqiue of visual imagery past and present, from paintings to pop videos. Semiotics is one tool among others which can be used to lay bare the contradictions and prejudices beneath the msmooth surface of the beautiful." [Iverson, 1986: p. 84]   Thus, semiotics, important to computer science generally is especially important for computer art. However, algorithmic art can answer the challenge that it feeds into the hierarchical and patriarchal rationali(sa)ti(ons) of antiquity and early modernity by pointing to interactive art! Interactive art allows algorithmic art to express not the dominance of the classical world view but the potential of the audience to become the artist.

### a. Structuralism

Structuralism was the operationalisation of reason in art filtered, to varying extents, through Marxism. Given that reason is orderly and that class structure permeates and defines society the structuralists sought to examine and expose the orderly system.  "Structuralism was generally satisfied if it could carve up a text into binary oppositions (high/low, light/dark nature/culkture and so on) and expose the logic of their working." [Iverson, 1986: p. 88.]

Structuralism grows out of the understanding that form and structure are really inseperable: "Meaning and artistic form are not easily separated in rerpesentation". [Iverson, 1986: p. 82.]

### b. Post Structuralism

Post-structuralists however expressed the Marxist sub-text more clearly: rather than merely exposing the structures of power, post-structuralists also seek to show how those structures are not in fact orderly and are rife with contradictions. "Deconstruction tries to show how such oppositions, in order to hold themselves inplace, are sometimes betrayed into inverting or collapsing themselves or need to banish to the text's margins certain niggling details which can be made to return and plague them." [Iverson, 1986: p. 88]

Post structuralist thought influenced semiotics not only by influencing theories of art but also by influencing linguistics. Saussure, and Peirce described an analytical approach to language. The former was static, the latter dynamic -- Saussure took "snapshots" of language, Peirce made "films" of language. Post-structuralists such as Chomsky describe a generative theory of linguistics. Structuralism permitted description of natural language "in terms of a system of opposing units that combine with each other in principled ways. However generative approaches (Chomsky, 1965) have been needed to account for th eunderlying process that make surface structures appear the way they do. The same applies to semiotics, as Eco shows." [deSouza, 1993: p. 762]. Just as this was true of linguistics it was also true of semiotics. "A taxonomy of signs contributes to describing semiotic systems but not to specifying the processes by which signs are used in communication.

Taxonomical approaches based on the icon-index-symbol distinction are by nature structure-oriented. They may support testing and evaluation of designed codes, but they fail to support the design process of metacommunication artefacts, since they do not explain or predict what happens when signs must be put together to create messages." [deSouza, 1993: p. 762]

B. Algorithm and Art

Algorithmic art is art generated mathematically. The algorithm, expressed as a computable function, can be implemented by a computer. However algorithmic art at first seems to be a self contradiction: Algorithms are about making rules. Art is often about breaking rules. How does this tension, between rule making and rule breaking resolve itself? According to Frieder Nake, the tension between art and algorithm is resolved by making artistic algorithms explicit. Nake describes several analogies within algorithmic art ("artificial art") and natural art (i.e. art which is not generated algorithmically by a computer).

1. Random Art

Most of Nake's early computer art used random number generators. (Nake: 2005) "My computer art programs soon developed to a point where they first had to decide which distribution function and which random number generator to use."(Nake: 2005) Randomness was a key element of all the early generative artists.

But randomness alone would be boring if not nonsensical. A random function does not carry much information at all beyond perhaps its periodicity and frequency. Since semiotics is about signification we have to look deeper in computer generated art to find where the signification is occurring in the communication of artist's message to an auditor and/or audience.

2. Painting by Constraints

In addition to randomness, Nake used constraints and problem solving within these constraints to generate his artwork:

> "A set of numerical constraints was specified that a picture construction was to satisfy. The generative aesthetics tried to solve the set of constraints. If that turned out to be possible, a picture was determined as a probability distribution over the set of permitted elemental signs. A quadtree construction process then dispersed the signs onto the picture." (Nake: 2005)

3. Analogies of machine art

The two main elements of Nake's art are randomness and problem solving within constraints. How does that play out in terms of artistic signification? Nake tries to find analogies between computer generated art ("artificial art") and art which is not computer generated ("natural art"). For example: Style is to art as program is to computer generated art.

Algorithmic art can be seen in terms of its style. "A style is like a language with an internal order and expressiveness admitting a varied intensity or delicacy of statement." [Iverson, 1986: p. 83]. The first analogy Nake makes is between the style adopted by a group of painters and a programming methodology adopted by algorithmic artists: "Almost all early experimenters in visual computer art made use of random numbers. ... If we were talking about natural art (the opposite of artificial art) we would interpret such an event as style: the common manner by a group of artists to draw or paint.

In our case, however, a pattern of a variety like 'random polygon' would be attributed to its simplicity. Irrespective of the details of the program, its results look much one like the other. This indicates that style and program may have some commonalities. ... The macro aesthetics of such drawings consist of two components: the overall geometry and the set of probability distributions." (Nake: 2005)

4. Intuition

As well as analogizing between artistic style of a school of art and programming style of a given computer language Nake also analogizes between intuition in natural art and randomness in computer generated art:

> "Is it far-fetched to draw some sort of analogy between the artist's intuitive decisions and the relializations of random processes, which - as such - are governed by propbability distributions? Such an analogy would definitely be far-fetched if we claimed that the artist's intutition was equal to a set of probability distributions. It would still be far-fetched (and thus wrong) if we claimed that the artist's decision processing was simulated by the random process. We could, however, justify saying that in place of intution in a human creative process was taken by an ordered set of probability distributions in a computer generated process."[Nake: 2005]

C. Dialectics of Art and Algorithm

Any artistic work can be understood dialectically: A work of art is the dynamic synthesis of opposed forces in tension. Of course, that subjective representation in the mind of the artist may be objectively wrong. It is nonetheless, even if the artist is unaware of it, a result of dialectics. Any number of contradictions, oppositions and dualities express themselves through the artistic work. The artistic work is the product of their resolution. This resolution of tension by synthesis is result of a dialectical process. "If we treat a situation dialectically, we identify its polar aspects and view their tension as the driving force for development."[Nake, Grabowski: 2002].

1. Dialectics of Art and Algorithm

Algorithms present rules to be applied to solve some problem. Art, in contrast, is ordinarily seen as a creative process of rule-breaking. This is the contradiction between art and algorithm.

Dialectically speaking competing theses  are seen as opposite principles standing in relative or absolute opposition to each other. "Art" and "algorithm" are clearly such opposite tensions. Art and algorithm are in tension with each other. "Art corrupts the algorithm, and calculation destroys the masterpiece"; [Nake, 2002] "computability can exist only where chaos is also welcome" [Nake, 2002]. The resolution of this tension occurs through dialectics which can be expressed as semiosis. Semiosis is a dynamic process of sign interpretation. But signs are not static: The meaning of signs changes with time, place and culture. Semiosis itself has an aesthetic: "unity in variety, order in chaos, surprise in expectation..." [Nake, 2002] The computer both interprets and presents signs and thus the artistic algorithm can express itself through the algorithmic sign.

3. Synthesis

How does this understanding of the role of randomness and constraint resolution as methods in computer generated art and as analogies to intuition and style in natural art play out in the dialectical question of art and algorithm? First, it is clear that these synthetic elements do serve to resolve the tension between art and algorithm. The algorithm becomes artistic precisely because of a combination (synthesis) of randomness and structure (constraints). This creation in turn resolves the contradiction between art and algorithm because it literally and significantly illustrates the commonality of art and algorithm through the illustration (art) and its programmatic counterpart (algorithm - signification). This dual signification (artistic and programmatic) also validates the thesis Nake presents -- that the algorithmic sign is a dual representation of code which can be read by machines (machine code) and humans (the GUI, for example, but also output on any media). The artistic aspect of the algorithm and the dialectical synthesis of the competing poles of art and algorithm explain why the algorithm is not merely an affirmation of hierarchical rationalism reflected in classicism or imposed by early modernity but rather presents the computer as a tool for the implementation of the expression of humanity through artistic creation by a dialectical synthesis. This escape from hegemonic rationalisation via dialectical synthesis clearly anchors the artistic algorithm in post-structuralist late modernity.

D. Implementation - Counterfeit Art

To try to illustrate a semiotic analysis of algorithmic art I chose to look at Georg Nees's drawing "Corridor".

Dialectically, this piece could be seen as a conflict  In this section of the inquiry we finally present answers to our main question: what can semiotics contribute to the analysis of art and algorithm?

Analysis of the Signs

We can analyze algorithmic art using the syntactic, semantic and pragmatic levels. We can analyse the user interface of interactive art by examining whether icons, indices, or symbols are used and which are the most effective.

For example, in "Doors" symbols were used - buttons with text. In "Nachmachung" dialog boxes are invoked for user input. The user input is confined to predefined choices. This is considered good design style. Since I want the user to learn by doing, to be surprised and to be curious I did not use indexical or iconic buttons. Rather I used symbols (words). Small pictures (miniature reproductions) of the products of the algorithms could have been used on the buttons to represent the algorithm. However I thought that words would be more likely to surprise and incite the curiousity of the reader. If a younger audience were the addressees of the art though then a good argument could be made for iconic representations of the algorithms in the buttons.

### 1. Syntactics of Corridor

At the syntactic level, i.e. the level of raw perception, Doors is a black and white line drawing. There are no colors. Further, all lines in Doors are straght - there are no attempts to represent curves at all. On either side of the drawing we see one continuous line which forms several rectangular (rhomboid) polygons.

### 2. Semantics of Corridor

At the bottom, we see several lines converging on what semantically must be seen as a vanishing point. At the top of the drawing we see several rectangles which we easily semantically interpret as cubes and boxes, i.e. as representations in two dimensions of three dimensional objects. Those rectangles at the very top of the page are longer than those which appear at the center. We also notice that the lines in the polygons on either side of the page also appear to converge toward a point slightly off center in the page - the vanishing point. Semantically it is obvious: we are looking at a two dimensional representation of three dimensional space using the classic figure of the vanishing point. Less obvious, but nevertheless semantically clear, is that the cubes and boxes at the very top of the picture are larger than those at the center of the picture to represent that they are closer to us. It seems fairly clear, semantically, that this is a corridor, even though the polygons on either side (the walls) are not joining either the cubes on the top of the paper (the ceiling) or the rows of lines at bottom (the floor). Interestingly, though the painting is titled "Corridor" - which hints at the word "door" there are no doors, at least not in front of us.

Dialectically, the semantics of Corridor are a dualistic conflict between the curve and the line - with the line winning an absolute victory. There are simply no curves in the work at all! We might say that this is a dialectics of absolute opposition, where either one or the other dialectical principle must win out, at least as conceived and implemented by Nees. This is not the only dialectical signification of Corridor. We can also see Corridor as a conflict between random art and perspective. There the results appear more ambiguous: both perspective and randomness are everywhere in the picture. This might be seen as an example of relative opposition. In cases of relative opposition dialectical synthesis is possible by resolving the two apparently opposite tendencies into some third whole which both incorporates and transcends each relative opposition. Yet, we know in the art of Frieder Nake randomness dominates and perspective appears rarely if at all. So this choice of portraying line and curve, of randomness and perspective, as dialectical oppositions is made not by society but by the artist. Yet, this is not an example of a subjective interpretant, but an objective referant. Anyone would agree that there are no curves in Corridor, so it is an objective fact and thus a referent not an interpretant.

### 3. Pragmatics

The final step in the semiotic analysis is the question of the pragmatics of the drawing. Is there a door behind us? Where does the corridor lead? Are there doors eventually before us? Why are there cubes on the ceilings? These pragmatic questions will be answered subjectively and differently by each user.

### 4. Algorithmics of Counterfeit Corridor

Algorithmically, the art is "semi" random. Clearly there are non-random constraints: there must be a vanishing point (determined randomly? Perhaps - for the vanishing point is not at the center of the page). All horizontal lines must converge on the vanishing point There are no curved lines.

### 5. Utility of the Exercise

I tried to reconstruct "Corridor" as well as a few other art works to see if I could understand some of the algorithm. This is why I noted that the vanishing point is in fact off center (in my copy it is not, but could be programmed to any point desired). Also the cubes not only are larger, they also are pointed in different directions depending on whether they are on the right or left side of the ceiling - something I overlooked, just like the off-center vanishing point until I tried to back-engineer an algorithm to generate a similar picture. The polygons on the walls appear to me to be in fact only one continuous line.

### 6. Limits of the Exercise

The most important limitation is the opacity of the original algorithm. I do not know if the ratio of randomness to perspective elements was somehow calculated. I only have one example of the art produced by the algorithm, so back engineering the art to get to the algorithm behind it is inevitably limited due to the limited sample size.

Constructing an algorithm to draw this picture did not of course answer the pragmatic questions (where does the corridor lead? where did we come from?). It did however force me to see some of the syntactic elements I might have overlooked. It also led me to discover some other interesting graphic functions and was in all events fun and thus a good example of constructivist learning - learning by doing, and having fun by learning, resulting in a greater appreciation of the object of study.

### Future Work

If the duality of art and algorithm is resolved through the algorithmic sign where does that leave us as artists or scientists? First, we can further explore these concepts. Nake has roughed out the contours of a sketch of a semiotics of algorithmic art and much work must to be done to complete  this representation. Such scientific cooperation would be in line with Nake's constructivism. Second, we can take from Nake's incisive analysis of the contradiction and its resolution through semiosis to synthesize the apparent contradiction into a whole that is greater than either of its parts as a source for inspiration in our own scientific and artistic work. By becoming more aware of our code as a part of our artwork - always the case of software art, and increasingly the case in other generative art as well - and by becoming more aware of the artistic aspects of our scientific inquiry and method and of the potential for scientificity at least in a formal descriptive sense through semiosis we can become both better artists and scientists. Nake has discovered a new field of semiological inquiry - and there is much productive work to be done.